\documentclass[aps,prc,preprint,groupedaddress]{revtex4-1}

\usepackage {graphicx}
\usepackage{units}
\usepackage{amssymb}
\usepackage{amsmath}
\usepackage{hyperref}

\begin{document}




\title{Deformed Flux Tubes Produce Azimuthal Anisotropy in Heavy Ion Collisions}


\author{H.J. Pirner}
\affiliation{Institute for Theoretical Physics, University of Heidelberg, Germany}

\author{K. Reygers}
\affiliation{Physikalisches Institut, University of Heidelberg, Germany}

\author{B.Z. Kopeliovich}
\affiliation{Departamento de F\'{\i}sica
Universidad T\'ecnica Federico Santa Mar\'{\i}a; and
\\
Centro Cient\'ifico-Tecnol\'ogico de Valpara\'iso;\\
Casilla 110-V, Valpara\'iso, Chile\\}

\begin{abstract}
We investigate the azimuthal anisotropy $v_2$ of particle production in nucleus-nucleus collisions in the maximum entropy approach. This necessitates two new phenomenological input parameters $\delta$ and $\lambda_2$ compared with integrated multiplicity distributions. The parameter $\delta$ describes the deformation of a flux tube and can be theoretically calculated in a bag model with a bag constant which depends on the density of surrounding flux tubes. The parameter $\lambda_2$ defines the anisotropy of the particle distribution in momentum space and can be connected to $\delta$ via the uncertainty relation. In this framework we compute the anisotropy $v_2$ as a function of centrality, transverse momentum and rapidity in qualitative agreement with LHC data.  
\end{abstract}



\maketitle



\section{Randomness and Order}
\label{sec:intro}

A wide consensus has been reached that nucleus-nucleus scattering allows one to study equilibrium thermodynamics of the quark-gluon plasmas at temperatures $T$ varying between 150\,MeV and 700\,MeV. Special emphasis has been devoted to the cross-over transition between the quark gluon plasma and the hadron resonance gas. The respective lattice calculations supplemented by hydrodynamic calculations of various observables, like the azimuthal asymmetry $v_2$, provide evidence of hydrodynamical flow of hadronic matter under the assumptions that the system arrives at local equilibrium very early after the collision. A historical overview of the success of this approach is presented in Ref.~\cite{Gale:2013da}. More recently also a highly anisotropic expansion with viscosity, worked out in Ref.~\cite{Ryblewski:2011aq}, has been shown to avoid the appearance of a negative longitudinal pressure.

The maximum entropy model we propose is an alternative approach. It emphasizes the phenomenological aspects of the reactions dynamics and is based on the imbalance between longitudinal and transverse motion in these high energy reactions. It agrees with the common wisdom that randomness is important to describe the momentum dependence of the inclusive and correlated cross sections. 
Randomness comes about because many low momentum partons/particles, radiating new QCD partons at a primordial stage, interact during the collision and evolve very differently in the longitudinal and transversal directions afterwards. Partons are best described by the Bjorken fractional light cone momenta $x$, and the transversal momenta. Consequently a random distribution must respect not only the mean energy pumped into the collision, like in a gas confined in a volume, but the conservation laws in longitudinal and transverse directions. Therefore we look for the most random distribution of partons compatible with two constraints, namely the produced transverse energy and the sum of the fractional momenta $x_i$ of all particles created in each hemisphere, which must amount to unity. The most random distribution with these constraints is the light cone plasma distribution. We expect this distribution to be reached at an intermediate time $t \sim (0.5-1)\,$fm/$c$ in the c.m.\ frame. 

The maximum entropy approach allows one to include these nonequilibrium features of the inclusive particle distribution by using two parameters, the effective transverse temperature $\lambda$ and the softness parameter $w$, instead of only one parameter, the temperature, as in the equilibrium distribution. The first parameter $\lambda$ comes from the constraint that the total transverse energy of the produced partons is fixed. The second parameter $w$ guarantees that the sum of all partonic light cone fractions is unity for forward and backward particles separately. To simplify we consider only gluons participating in the collision, then the maximum-entropy method yields a Bose-type distribution depending on light cone fraction $x$ and transverse momentum as follows \cite{Pirner:2011ab}:
\begin{equation}
n(x,\vec p_{\bot})=
\frac{1}{e^{\frac{|\vec p_{\bot}|}{\lambda}+x w }-1}
\end{equation}
with
\begin{equation}
x= \frac{\epsilon+p_{z}}{E+P_{z}}. \nonumber
\end{equation}

In pp collisions the transverse configuration space is homogeneously distributed over the area $ L_{\bot}^2$. 
Inclusive cross sections give a size $ L_{\bot} \approx 1$\,fm for gluon distributions, cf. Ref.~\cite {Pirner:2011ab}.
In nucleus-nucleus collisions  we  add up the individual contributions of all nucleonic participants $N_\mathrm{part}$. 
As we have shown in Ref.~\cite{Pirner:2012yy}, the effective  transverse temperature $\lambda$ rises with centrality because of collisional broadening of the partons. Finally invoking parton-hadron duality we obtain the multiplicity $N/2$ of produced hadrons in each hemisphere of the cm-system  by integrating the light cone distribution over the respective phase space. Note \cite{Pirner:2011ab} that the relativistic measure $dx/x$ arises from the large spatial extension in longitudinal direction of the small $x$ partons and the gluon degeneracy factor is given as $g = 2 (N_c^2-1)$:
\begin{equation}
N/2=g \frac{N_\mathrm{part}}{2} L_{\bot}^2\int \frac{d^2 p_{\bot}}{(2 \pi)^2} \int \frac{dx}{x} n(x,\vec p_{\bot}).
\label{eq:mult}
\end{equation}

We emphasize that a statistical understanding of the final state in heavy-ion collisions necessitates both a correct description of the momentum and configuration space distribution.  With this in mind it is clear that changes must be implemented to describe the azimuthal anisotropy in the maximum-entropy method. One has to take into account that phase space is deformed for non-central collisions. This means that both configuration space and momentum space are not isotropic. The impact parameter $b$ along the $x$ axis and the momenta of the incoming particles along the $z$ axis determine the reaction plane. Orthogonal to this $x,z$ system lies the third $y$ axis. Let the angle $\varphi$ in the $x,y$ plane be the angle between the transverse momentum of the particle and the $x$ axis. Any asymmetry in transverse momentum space relative to the reaction plane has to be considered together with the deformed asymmetry in configuration space  given by the overlap zone of the two nuclei. There are indications from HBT measurements for a noticeable asymmetry in configuration space even much later in the collision cf.~\cite {Adams:2003ra}. 

\begin{figure}
\centering\includegraphics[width=0.5\linewidth]{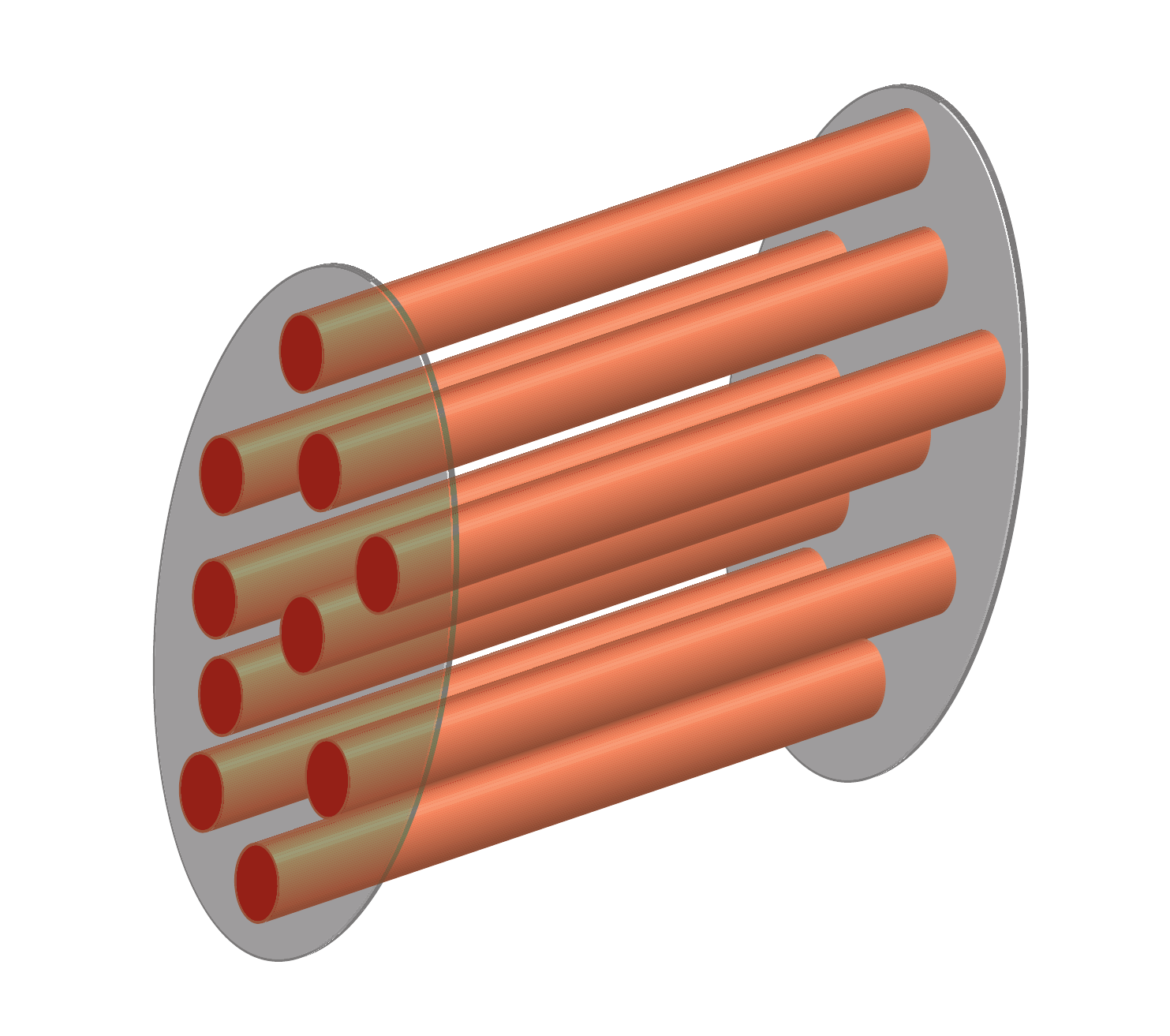}
\caption{Schematic picture of the flux tubes formed in the Glasma are shown for a non-central collision. A deviations from a circular transverse shape of a flux tube leads to anisotropic particle production.}
\label{fig:glasma}
\end{figure}

In the following, we concentrate on the early stage of the collision and consider the different manifestations of gluonic QCD degrees of freedom at very high energies. Shortly after the collision the scattered nuclei pull many color-flux tubes between the activated, color charged partons \cite{Casher:1978wy}. These tubes are gluonic degrees of freedom at low resolution. The transverse size and position of a tube fluctuate, and such fluctuations are certainly affected by presence of neighboring tubes. The tubes keep decaying, i.e. they break up due the Schwinger phenomenon and produce $\bar qq$ pairs.  Flux tube configurations are important in the initial state of the collision as e.g. discussed in the Glasma model \cite{Gale:2012rq}, c.f.~Fig.~\ref{fig:glasma}. Kinks can be associated with hard gluons as in the Lund model \cite {Andersson:2001tg}. 

The transverse momentum distribution of partons created in the homogeneous field of a flux tube is related to the area of the flux tube $L_\bot^2$ as \cite{Casher:1978wy},
\begin{equation}
\frac{dn}{d^2p_\bot} \propto e^{- \alpha \, p_\bot^2 L_\bot^2} \,.
\label{eq:flux}
\end{equation}
The energy of the color flux per unit of length (string tension) for tubes formed between color-triplet charges is related to the slope of meson Regge trajectories, $\kappa=1/(2\pi\alpha_{\mathbb{R}}^\prime)\approx 1$\,GeV/fm. The string tension of an color octet-octet tube is expected to be much larger, because the slope parameter of the Pomeron trajectory, $\alpha_{\mathbb{P}}^\prime \approx 0.25\,{\rm GeV^{-2}}$ is at least 4 times smaller. Data from HERA prefer even smaller $\alpha_{\mathbb{P}}^\prime\approx 0.1\,{\rm GeV^{-2}}$ \cite{Kopeliovich:2000pc}. Note that a part of the observed elastic slope comes from unitarity saturation\cite{Kopeliovich:2000ef}. As far as the energy of two triplet-antitriplet strings is lower than the energy of a single octet-octet string, we conclude that $ 3 \bar 3$ flux tubes repel each other at short transverse distances. The interaction of QCD flux tubes certainly depends on the distance of the centers of the tubes, and it is possible that their long range interaction \cite {Kalaydzhyan:2014zqa} is attractive before the repulsion sets in. 

QCD flux tubes resemble vortex lines in type II superconductors \cite{Bicudo:2010gv}. Such vortices can also be studied in a Bose-Einstein condensate of cold atoms and have been simulated in an anisotropic trap in Ref.~\cite{McEndoo:2009}. We conjecture that the flux tube environment in heavy-ion collisions will also influence the transverse shape of QCD flux tubes. The presence of many surrounding flux tubes dilutes the QCD vacuum, decreasing its energy density, which leads to a reduction of the bag constant, i.e., to a reduction of the vacuum pressure on the bag walls. The tube is swelling more in the direction where the density is higher. Therefore, we parametrize the available configuration space for the gluons by two different extensions $L_x$ and $L_y$ in the $x$ and $y$ directions. One may think that these extensions are related to the major and minor axis of the ellipsoidal area of the flux tube. 

The transverse tube radius certainly increases, but for the sake of simplicity we keep the average cross section of a flux tube fixed and use it is as a parameter which has been taken from previous fits to the inclusive cross section (cf.\ Eq.\ (\ref{eq:mult})):
\begin{equation}
L_x L_y=L_{\bot}^2.
\end{equation}
The configuration space asymmetry is described by the a parameter
\begin{equation}
\delta =\frac{L_x^2-L_y^2}{L_x^2+L_y^2}.
\end{equation}
We define $\bar \delta(b)$ as the average of $\delta$ over all flux tube created in a heavy-ion collision at impact parameter $b$.
Note that the parameter $\bar \delta$ differs from the parameter $\varepsilon$ given in the literature, e.g.\ \cite{Heinz:2013th}, which represents the average of $\langle x^2-y^2\rangle/\langle x^2+y^2\rangle$ of the overlap area in the nucleus-nucleus collision.

The observed anisotropy in momentum space necessitates a third constraint in the  maximum-entropy method. It appears natural to use the second azimuthal moment of the total transverse energy as an additional constraint. Then the complete set of constraints is the following:
\begin{eqnarray}
g  L_x L_y\int \frac{d^2 p_{\bot}}{(2 \pi)^2} \int \frac{dx}{x} 
x  n(x,\vec p_{\bot},\varphi)&=&1\\
g  L_x L_y\int \frac{d^2 p_{\bot}}{(2 \pi)^2} \int \frac{dx}{x} 
|\vec p_{\bot}| \ n(x,\vec p_{\bot},\varphi)&=& \langle E_{\bot,pp} \rangle\\
g  L_x L_y \int \frac{d^2 p_{\bot}}{(2 \pi)^2} \int \frac{dx}{x} 
|\vec p_{\bot}| \cos[2 \varphi] n(x,\vec p_{\bot},\varphi)&=& c_2 \langle E_{\bot,pp} \rangle.
\end{eqnarray}
These three constraints are added to the entropy per participant with three Lagrange parameters. Then the sum is maximized: 
\begin{equation}
\frac{\delta \Big( S+\frac{1}{\lambda_0} \sum |p_{\bot}| n(x,p_{\bot},\varphi)+ w \sum x  n(x,p_{\bot},\varphi)+\frac{1}{\lambda_2}\sum |p_{\bot}| \cos[2 \varphi] n(x, p_{\bot},\varphi) \Big) }{\delta n(x,p_{\bot},\varphi)}=0 \, .
\end{equation}

Since all constraints are linear, the resulting maximum entropy distribution has the familiar exponential form of the light-cone Bose distribution. The third constraint defines a third Lagrange parameter $\lambda_2$, the transverse energy asymmetry. The distribution takes the form
\begin{equation}
n(x,p_{\bot},\varphi)=
\frac{1}{e^{|\vec p_{\bot}|(\frac{1}{\lambda_0}+\frac{\cos[2 \varphi]}{\lambda_2})+x w }-1} \, .
\label{eq:lcp}
\end{equation}

Higher moments  on the transverse energy would generate more parameters. Fluctuating distributions of nucleons in the nuclei can produce odd moments. These more general cases needing extra Lagrange parameters can be treated in a similar way as above. The strictly phenomenological maximum entropy method tries first to parametrize the data without prejudices concerning the reaction mechanism. In a second and separate step this approach aims to understand the parameters entering this description. We think this separation is helpful to get the physics underlying 
the QCD plasma correctly. Restricting ourselves to the second azimuthal moment, we get the following parton distribution: 
\begin{equation}
\frac{d N_{AA}}{dy p_{\bot}dp_{\bot} d\varphi}= g \frac{N_\mathrm{part}}{2}\frac{ L_x L_y} { (2 \pi)^2} \frac{1}
{e^{|\vec p_{\bot}|(\frac{1}{\lambda_0}+\frac{\cos[2 \varphi]}{\lambda_2}+\frac{ w e^{|y|}}{\sqrt{s}})}-1}
\end{equation}

If parton-hadron duality holds, this  generalized maximum-entropy distribution gives the semi-inclusive cross section of hadrons in AA-collisions including azimuthal asymmetry. We have seen in previous publications \cite{Pirner:2011ab,Pirner:2012yy} that duality holds approximately. Hadronic masses modify the light-cone distribution, probably also the azimuthal asymmetry. Concentrating on the central rapidity window ($y \approx 0,\frac{w}{\sqrt{s}}\ll \frac{1}{\lambda_0}$) one can integrate the above inclusive cross section with the weight function $\cos(2 \varphi)$. Representing the Bose distribution by a geometric series one obtains a simple expression for the total asymmetry $v_2$, since  $|\lambda_2|\gg \lambda_0$:
\begin{equation}
v_2=\frac{\int \frac { d N_{AA}}{dy p_{\bot}dp_{\bot} d\varphi}|_{y=0} \, p_{\bot} \cos(2 \varphi) \, d \varphi d p_{\bot}}{
              \int \frac{d N_{AA}}{dy p_{\bot} dp_{\bot} d\varphi}|_{y=0} \, p_{\bot} \, d \varphi d p_{\bot}}\\
   \approx -\frac{\lambda_0}{\lambda_2}.           
\end{equation}
A small value of $v_2$ points to an hierarchy of scales manifested in $\lambda_2$ being much larger than $\lambda_0$. This hierarchy also reflects that flux tubes do not fragment independently from one another, but do interact slightly. 

\section{Geometry of the collision and flux tube deformation}
In the following we assume a simple bag-model \cite{Chodos:1974je} for the flux tube and its support through QCD sum rules \cite{Shifman:1978by}. For heavy ion collisions, it is important to consider that the flux tube is inserted into an environment of nearby flux tubes. The number of flux tubes is proportional to the density of participants $N_\mathrm{part}(\vec x_{\bot})$. The higher this density the more the flux tube properties will be affected. The bag constant $B$ in the environment will differ from the bag constant $B_0$ in vacuum. This approach is in the spirit of models in which nucleon radii in nuclei depend on the surrounding density \cite{Close:1984zn}. We assume that  the bag pressure decreases linearly with the density and deconfines the partons, once the density of participants approaches a critical value $n_0$ (B $\equiv$ 0 for $n \geq n_0$):
\begin{equation}
B(N_\mathrm{part}) =B_0 (1-\frac{N_\mathrm{part}(\vec x_{\bot})}{n_0}) \, .
\end{equation}

We expand the local bag constant in the transverse overlap plane around $\vec x_0$ up to second order in distance:
\begin{eqnarray}
B( \vec x)&\approx &B (\vec x_0) +x \frac {\partial B}{\partial x}|_{\vec x_0} +y \frac {\partial B}{\partial y}|_{\vec x_0}+...\\
          &=&B (\vec x_0) +x B_x|_{\vec x_0} +y B_y|_{\vec x_0} \, .
\end{eqnarray}
The longitudinal direction of the flux tube remains homogeneous in longitudinal direction, so its energy per length gives the string tension $\kappa$.
If $Q$ is the external charge at the ends of the flux tube, then we get the string tension $\kappa$ by integrating over the area  $L_x L_y$ of the flux tube:
\begin{equation}
\kappa=\frac{1}{2} \frac{Q^2}{L_x L_y} + B (\vec x_0) L_x L_y +\frac{1}{2}B_x L_x^2 L_y  +\frac{1}{2} B_y L_y^2 L_x.
\end{equation}
This string tension has to be minimized under the constraint that the total area of the flux tube is unchanged:
\begin{equation}
L_x L_y= L_{\bot}^2 \, .
\end{equation}

The solution depends on the sign of the gradients of the bag pressure. Since on the left side of the mandola formed in the overlap of the nucleus-nucleus collision the gradients are opposite to the right side, we can combine the two sides by taking the absolute values of the gradients. In agreement with intuitive expectation we find that the flux tube is deformed when the local pressure is larger on one side of the bag than on the other, quantitatively the deformation depends on the gradients of the bag constant in the $x$ and $y$ directions: 
\begin{eqnarray}
L_x^2= \left|\frac{B_y}{B_x}\right| L_{\bot}^2\\
L_y^2= \left|\frac{B_x}{B_y}\right| L_{\bot}^2 .
\end{eqnarray}
Using in these equations the density dependent bag constant we calculate how the relative extensions $L_x$ and $L_y$ vary inside the overlap region of the nucleus-nucleus collision for each impact parameter $b$. The strongest differences between $L_x$ and $L_y$ occur at the edges of the overlap regions. In the following $L_x$ and $L_y$ will always be denoting the mean value of these extensions averaged over the whole area. Obviously the first order expansion breaks down when $B_x$ or $B_y$ are equal to zero at the $y$ or $x$ axis, but these lines are not important for the averages. The next section will show how the deformation of configuration space will affect the momentum distribution.

\section{Geometry of the collision and momentum asymmetry}
On first sight, cf. Eqs.~ (11) and (12), the configuration space asymmetry $L_x \neq L_y$ cancels out in the ratio for $v_2$. But we should not leave out, how quantum mechanics generates a momentum anisotropy for the gluons inside the flux tubes when the flux tubes are deformed. 
In Ref.~\cite{Sailer:1990ia} the momentum distribution of hadronization fragments has been calculated for spherical flux tubes. In our dual picture we do not differentiate between gluons and hadrons. Recalling the relation between the size of the flux tube and the momenta of partons  given in Eq.~\ref{eq:flux} we estimate the momentum asymmetry of the gluons from the uncertainty principle: 
\begin{eqnarray}
\frac{\langle p_x^2-p_y^2 \rangle}{\langle p_x^2+p_y^2 \rangle}&=&\frac{\frac{1}{L_x^2}-\frac{1}{L_y^2}}{\frac{1}{L_x^2}+\frac{1}{L_y^2}}\\
                                   &=& - \delta \, .
\end{eqnarray}
We can also calculate this asymmetry in leading order $\lambda_0 / \lambda_2$ from the deformed light-cone plasma distribution of gluons  given in Eq.~(\ref{eq:lcp}):
\begin{equation} 
\frac{\langle p_x^2-p_y^2 \rangle}{\langle p_x^2+p_y^2 \rangle} =- \frac{2 \lambda_0}{\lambda_2}.
\end{equation}
Combining the two equations we are able to relate the momentum space asymmetry $\lambda_2$ to the configuration space anisotropy $\delta$. This is the main result of this section:
\begin{equation} 
\lambda_2= \frac{2 \lambda_0}{\delta}.
\end{equation}

Finally, we compute the momentum integrated $v_2$  from the anisotropic light-cone plasma
distribution with the parameter $\lambda_2$  (Eq.~(12)). The resulting $v_2$ is simply related  to the 
geometrical function $\bar \delta(b)$ which we calculate from the overlap area averaged asymmetry. Note that in this derivation the parametrization of the light-cone distribution with the parameters $\lambda_0$ and  $\lambda_2$ enters:
\begin{equation}
v_2(b)  \approx -\frac{\bar \delta(b)}{2}.      
\end{equation}

\begin{figure}
\centering
\includegraphics[width=0.95\linewidth]{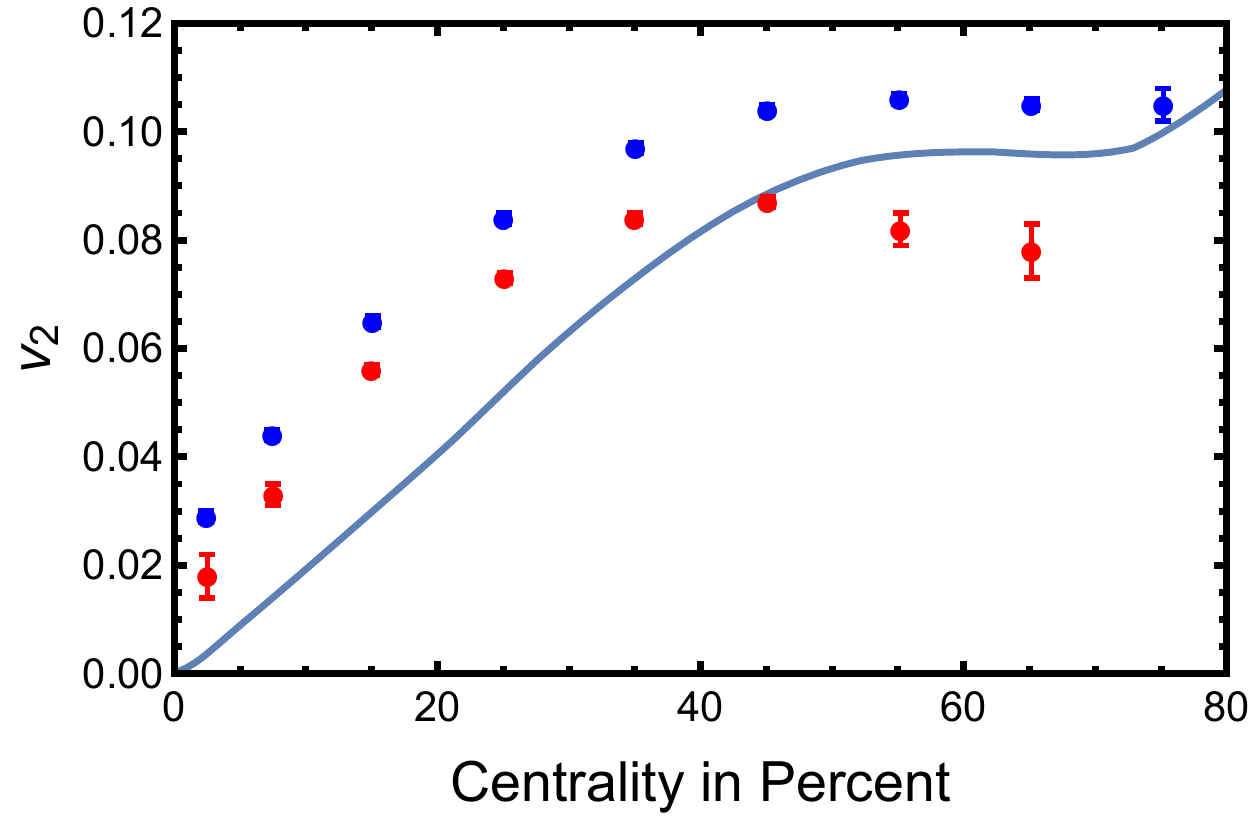}
\caption{The theoretical azimuthal asymmetry $v_2$ is shown as a function of centrality in percent. The data points represent measurements of $v_2\{2\}$ (blue points) and $v_2\{4\}$ (red points) for charged particles from ALICE \cite{Aamodt:2010pa} at $\sqrt{s_{NN}}=2.76$ TeV.}
\label{fig:v2percentage}
\end{figure}

In Fig.~\ref{fig:v2percentage} we compare the so calculated azimuthal asymmetry $v_2$ for massless gluons  with the data for $v_2\{2\}$ (blue points) and $v_2\{4\}$ (red points) of charged particles as a function of the centrality for Pb-Pb collisions at LHC at $\sqrt{s_{NN}}=2.76$ TeV. We reproduce the general behavior of $v_2$, but we underestimate the data  for low centralities where fluctuations 
in the positions of the participants are important. Integrating over all impact parameters one obtains an average value $\lambda_2$ for  LHC: 
\begin{eqnarray} 
\lambda_2 &=&\unit[-6.9]{GeV}. \nonumber
\end{eqnarray}

Negative $\lambda_2$ gives a positive value for $v_2(p_{\bot})$. The large  absolute value of the transverse asymmetry energy $\lambda_2$ compared with the  transverse effective temperature $\lambda_0 \approx 0.38$\,GeV  generates a small anisotropy. 

The presented model is based on the early stage of the collision, where the flux tube geometry plays an important role. As noted in Ref.~\cite{Heinz:2013th}, any sort of interaction among the primordial QCD degrees of freedom will cause non-vanishing radial and anisotropic flows even before the system has thermalized and viscous hydrodynamics becomes applicable. Of course strong interactions of the finally produced hadrons may also modify the angular anisotropy. Recently another model of the initial state interaction based on AdS/CFT \cite{Shuryak:2013sra} has come to a different conclusion about the deformation of the flux tubes. It proposes a widening of the flux tubes along the impact parameter axis. Since AdS/CFT has special problems to describe QCD in the region between $T_c$ and $3 T_c$ \cite{Megias:2010ku,Veschgini:2010ws} we think its predictive power for such complicated processes is limited. The simple bag model may do better.

A momentum dependent $v_2(p_{\bot})$ can be computed from the azimuthal integrals of the anisotropic light-cone plasma distribution which give sums of modified Bessel-functions $I_1$ and $I_0$ of the first kind
\begin{eqnarray}
v_2(p_{\bot})&=&\frac{\int \frac { d N_{AA}}{dy p_{\bot}dp_{\bot} d\varphi}|_{y=0} \, \cos(2 \varphi) d \varphi}{
\int \frac{d N_{AA}}{dy p_{\bot} dp_{\bot} d\varphi}|_{y=0} \, d \varphi}\\
 \nonumber
&=&\frac{\sum_{n=1}^{\infty} (-1) \exp(- n p_{\bot}/\lambda_0) I_1(\frac{n p_{\bot} }{\lambda_2})}{\sum_{n=1}^{\infty} \exp(- n p_{\bot}/\lambda_0) I_0(\frac{ n p_{\bot}}{\lambda_2})}.
\end{eqnarray}

For the LHC at $\sqrt{s_{NN}}=2.76$\,TeV we have the parameters given in Table~\ref{tab:parameters}. These parameters are fitted at each centrality bin to the constraints $dN/dy$ at $y=0$, $\langle p_ {\bot} \rangle$ and the $x$-sum rule. The effective transverse temperatures and the transverse size for massless gluons differ slightly from the corresponding parameters for massive pions. The effective temperature $\lambda_0$ rises for more central collisions due to parton rescattering as explained in Ref.~\cite {Pirner:2012yy}. The asymmetry energy $\lambda_2$ increases much more strongly with centrality. The transverse size is constant  $L_{\bot}=0.766$\,fm. 
\begin{table}[h]
\begin{tabular}{c|c|c|c}
Centrality & $\lambda_0$ (GeV) & $\lambda_2$ (GeV) & $w$ \\
\hline
10-20\% & 0.396 & -9.25 & 7.2 \\
20-30\% & 0.383 & -5.65 & 6.73 \\
30-40\% & 0.376 & -4.48 & 6.49 
\end{tabular}
\caption{Parameters of the anisotropic light-cone distribution for different centralities in Pb-Pb collisions at $\sqrt{s_\mathrm{NN}} = 2.76$~TeV.}
\label{tab:parameters}
\end{table}

\begin{figure}[ht]
\centering
\includegraphics[width=0.95\linewidth]{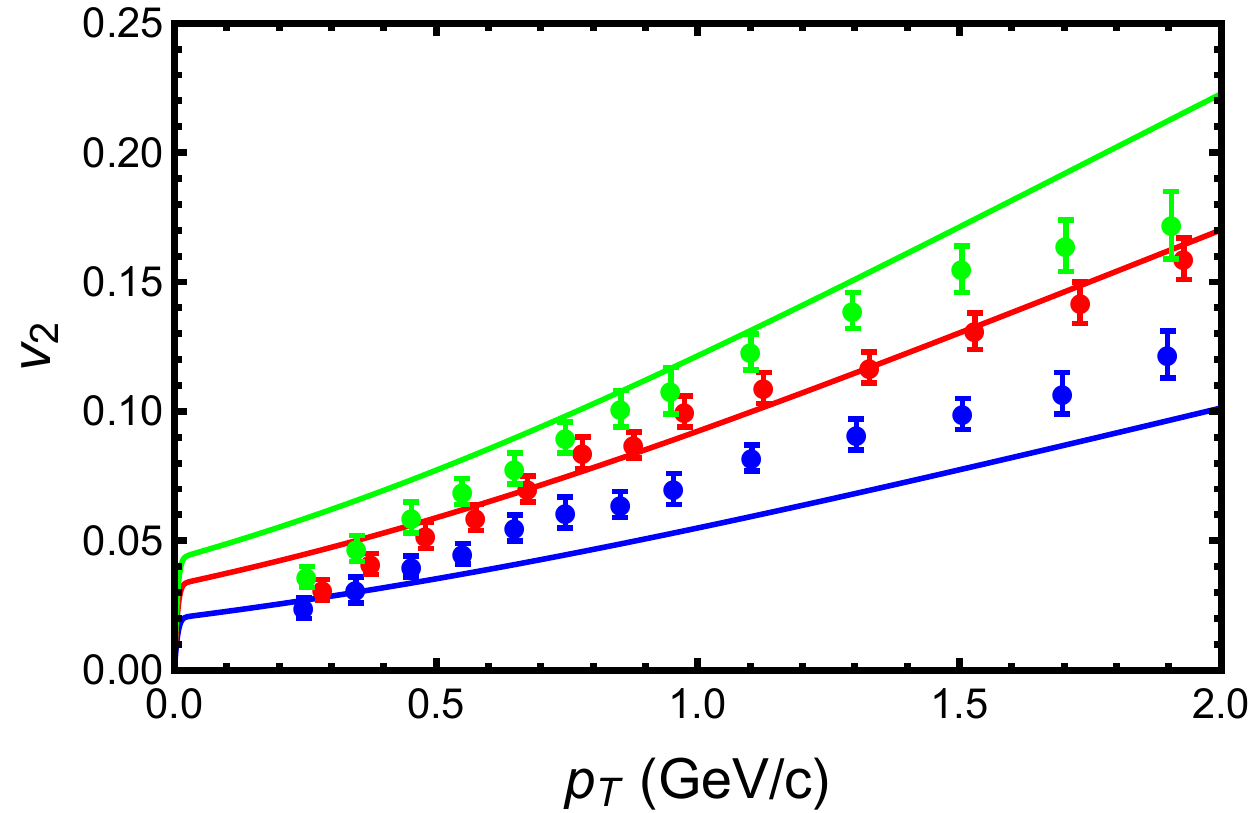}
\caption{The azimuthal asymmetry $v_2$ is shown as a function of transverse momentum, in three different centrality bins $(30-40\%, 20-30\%,10-20\%)$ from top to bottom for LHC at $\sqrt{s_{NN}}=2.76$\,TeV. The data points are $v_2\{4\}$ measurements from ALICE \cite{Aamodt:2010pa}.}
\label{fig:v2gluons}
\end{figure}

In Fig.~\ref{fig:v2gluons} we show the resulting momentum dependent asymmetries
for LHC at $\sqrt{s_{NN}}=2.76$\,TeV in these three different centrality bins. The different slopes of $v_2$  as
a function of transverse momentum are reproduced. With masses of the hadrons included in the constraints of the maximum entropy distribution the anisotropy will be suppressed for small momenta. 
Since $|\lambda_2|>>\lambda_0$, we can expand the Bessel functions for small transverse momentum $p_{\bot}\rightarrow 0$ and obtain a formula which is good for $0.1 < p_{\bot} <1$\,GeV. For small momenta the slope is given by the inverse of the transverse asymmetry energy. Therefore, peripheral collisions have steeper slopes.

\begin{equation}
v_2(p_{\bot})=-\frac{1}{\lambda_2} (\frac{\lambda_0}{2}+\frac{p_{\bot}}{4}+\frac{p_{\bot}^2}{24 \lambda_0}) .
\end{equation}
Since we have the explicit form of the rapidity dependence of the light-cone plasma distribution, we can also calculate  the momentum integrated flow parameter as a function of the  rapidity $y$ with the parameters of Table~\ref{tab:parameters} used for the calculation of the momentum dependence and $K=0.35$ given in Ref.~\cite{Pirner:2012yy}
\begin{equation}
v_2(y)=-\frac{1}{\lambda_2} (\frac{1}{\frac{1}{\lambda_0} +\frac{ w e^y}{K \sqrt{s}}}).
\end{equation}

\begin{figure}[ht]
\centering
\includegraphics[width=0.95\linewidth]{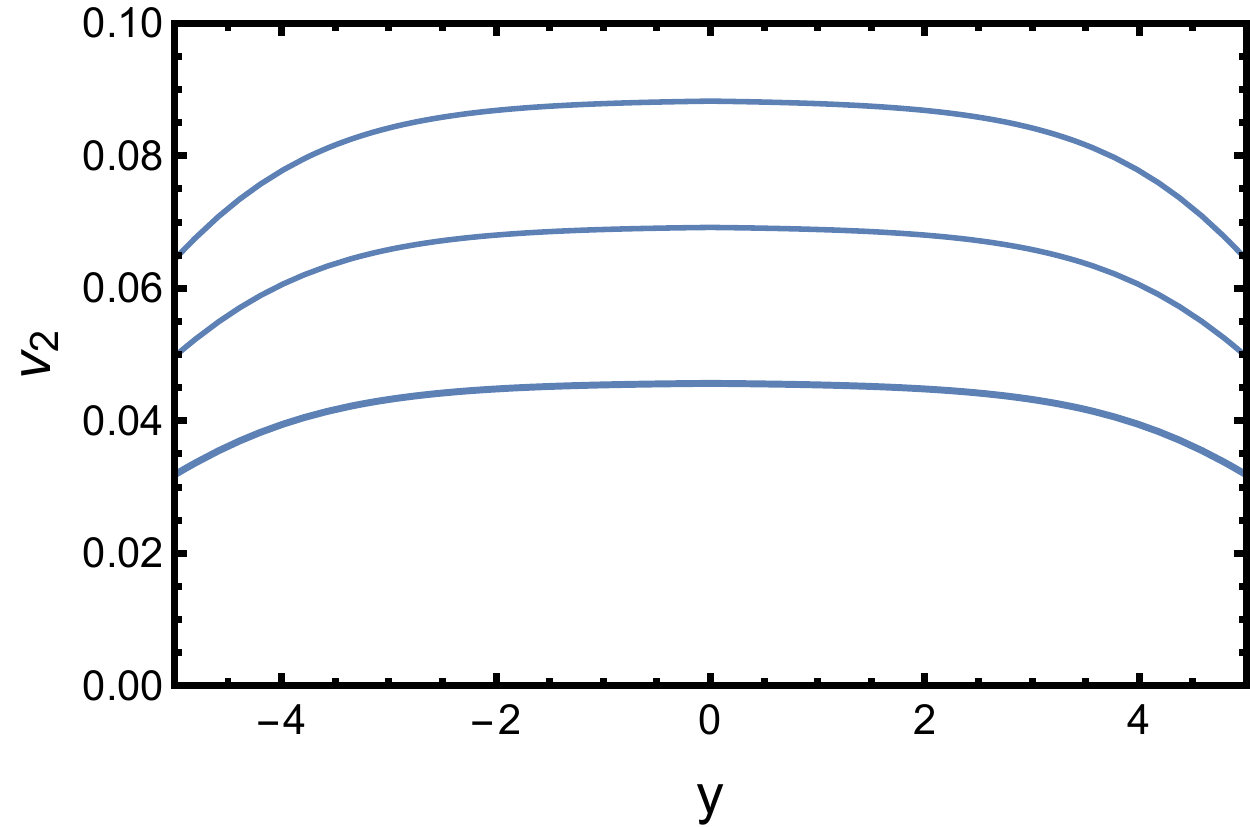}
\caption{The impact parameter integrated azimuthal asymmetries $v_2$ are shown as a function of rapidity using the deformed light-cone plasma distribution for centrality bins $(30-40\%, 20-30\%,10-20\%)$ from top to bottom.}
\label{fig:v2y}
\end{figure}
The anisotropy $v_2(y)$ depends on rapidity in the above specific combination including the effective transverse temperature $\lambda_0$, the softness
parameter $w$ and  the effective cm-energy $K \sqrt{s}$. In Fig.~\ref{fig:v2y} we show the resulting rapidity dependent asymmetries. For each centrality there is only one calculated $\lambda_2$ in the light-cone plasma distribution to describe the dependence of the anisotropy parameter $v_2$ on momentum and rapidity. This is  a definite advantage of the light-cone plasma distribution compared with other statistical distributions. The theory at $y=0$ agrees with the data points displayed in Fig.~\ref{fig:v2percentage}. 

\section{Discussion}
We have presented a model for the anisotropy parameter $v_2$ based on the non-equilibrium maximum-entropy distribution. We argue that deformed gluon flux tubes cause the momentum anisotropy $v_2$ which develops at an intermediate stage of the collision described by the non-equilibrium gluon light-cone plasma. From the momentum scale $\lambda_0$ of the light-cone distribution one can estimate this time to be $\tau \approx \frac{1}{\lambda_0} = 0.5$\,fm/$c$. 

In various models of heavy ion collisions, like the EPOS model \cite{Pierog:2013ria} and the Glasma picture \cite{Gale:2012rq}, flux tubes play an important role, and one must investigate their properties. These should be compatible with an increasing tendency to deconfine as seen in a thermal environment at high temperature. With a bag constant which depends linearly on the density of participants, one can quantify how the flux tube behaves in the environment. From this calculation a deformation results corresponding to the different gradients of the participant densities in different directions. We remark that these gradients also played a role in a perturbative calculation of the azimuthal anisotropy of photons \cite{Kopeliovich:2007fv} which however could not explain the size of the observed effect. In our model the total cross section $L_{\bot}^2$ of the flux tube is constrained from the integrated multiplicity and was kept constant. It cannot account for very high densities where the environment would also affect the total area which has to be studied separately. Especially high densities are investigated in a recent article \cite{Kalaydzhyan:2014zqa} where an "implosion" of strongly overlapping flux tubes is discussed. 
 
We emphasize the quantum properties of the flux tube. Quantum mechanically, the flux tube size gives the right size of transverse momenta. The uncertainty relation connects the spatial asymmetry $\delta$ to the momentum asymmetry of gluons. This way, we can determine the transverse asymmetry energy $\lambda_2$, the third parameter in the maximum entropy distribution. 
 
The presented paper restricts itself to the asymmetry of the gluons, therefore it presents only a crude picture of the full collisions dynamics. Further work is necessary to investigate additional consequences following from our picture. For instance, deformed flux tubes can be used as a model for pre-equilibrium flow to initialize a further hydrodynamic evolution. On a purely phenomenological level it is also worthwhile to determine the transverse asymmetry energy parameter $\lambda_2$ in Eq.~(8) directly from the experimental data. The more facets we learn about high energy $pp$ and nucleus-nucleus collisions the more the early stages of the collision seem to become important reflecting the light-like trajectories of the partons and their dynamics.   
  
\begin{acknowledgments}

This work was supported in part
by Fondecyt (Chile) grant 1130543.
\end{acknowledgments}





\bibliography{deformed_flux_tubes}






\end{document}